\documentclass[showpacs,preprintnumbers,nofootinbib,
               amsmath,amssymb,superscriptaddress]{revtex4}

\newcommand{\feyn}[1]{
  \setbox0=\hbox{\ensuremath{#1}}
  \hbox to\wd0{\hbox to0pt{\hbox to\wd0{\hss/\hss}\hss}\box0}}
\newcommand{\x}{\text{x}}

\newcommand{\tr}{\text{tr}}
\newcommand{\A}{\mathcal{A}}
\newcommand{\D}{\mathcal{D}}
\newcommand{\K}{\mathcal{K}}
\newcommand{\N}{\mathcal{N}}
\newcommand{\W}{\mathcal{W}}
\newcommand{\dA}{\delta A}
\newcommand{\brho}{\bar{\rho}}
\newcommand{\drho}{\delta\brho}
\newcommand{\Sym}{S_{\text{YM}}}
\newcommand{\xt}{\boldsymbol{x}_\perp}
\newcommand{\yt}{\boldsymbol{y}_\perp}
\newcommand{\zt}{\boldsymbol{z}_\perp}
\newcommand{\partt}{\boldsymbol{\partial}_\perp}

\begin{document}

\title{Deriving the Jalilian-Marian-Iancu-McLerran-Weigert-Leonidov-Kovner
       equation with classical and quantum source terms}
\author{Kenji Fukushima}
\affiliation{RIKEN BNL Research Center, Brookhaven National
             Laboratory, Upton, New York 11973-5000, USA}
\begin{abstract}
 We discuss the high-energy evolution equation in Quantum
 Chromodynamics (QCD) in the framework of the color glass condensate.
 We rewrite the generating functional of QCD with the color glass
 condensate into a representation with the density of states and a
 simple eikonal coupling.  In this representation we introduce an
 auxiliary variable which is identified as the quantum color charge
 density consistent with non-commutativity in the operator formalism.
 We revisit the derivation of the
 Jalilian-Marian-Iancu-McLerran-Weigert-Leonidov-Kovner (JIMWLK)
 equation to clarify how the non-commutative nature among the color
 charge density operators can be disregarded in the regime where the
 gluon density is high.
\end{abstract}
\preprint{RBRC-587}
\maketitle


\section{introduction}

     There have been remarkable developments in our understanding of
high-energy description of Quantum Chromodynamics (QCD) in the last
decade.  Although, in principle, main constituents of nuclei could be
thoroughly governed by QCD dynamics, it would require enormous
numerical computation to perform a first-principle calculation of QCD
for the properties of nuclei.  Most of problems encountered in QCD are
non-perturbative, and besides, non-linear, meaning that gluons
strongly interact with themselves.

     Under some conditions feasible in experiments, however, one may
hope to derive analytical information directly from QCD in a
manageable way, which is not only of general interest in theoretical
physics, but also of importance as an experimental testing ground for
theoretical ideas in QCD.\ \ It is well known that the strong coupling
constant, $\alpha_s$, is expected to be small for far energetic
processes owing to asymptotic freedom, so that the perturbation theory
would work.  In fact, however, the perturbative expansion becomes
reliable for processes in which the exchanged gluon carries a large
momentum and it is possible that the momentum transfer remains small
even for energetic processes.  Moreover, even if the exchanged gluon
conveys a sufficiently large momentum transfer, small $\alpha_s$ is
not necessarily adequate to guarantee the validity of perturbative
calculations.

     Even when the perturbative expansion in terms of $\alpha_s$ seems
plausible, resummation is needed whenever enhancing logarithms exist
in the expansion series, which occurs in general for processes
characterized by multiple (large) energy scales.  This is actually the
case for the processes we are discussing in this paper in which
partons with small Bjorken-$\x$ are relevant.  Then, two large scales
$\Lambda_{\text{QCD}}\ll t\ll s$ may give rise to a large logarithmic
factor $\sim\ln[s/t]\sim\ln[1/\x]$.  Here Bjorken-x is intuitively
interpreted as the longitudinal momentum fraction $\x=p^+/P^+$ of a
parton inside a target seen in the infinite momentum frame,
$P^+\to\infty$;  the target moves nearly at the speed of light in the
positive $z$ direction, that is, along the $x^+=(z+t)/\sqrt{2}$
direction in the light-cone coordinate.  Even though $\alpha_s$ is far
smaller than unity, the perturbation theory in $\alpha_s$ breaks down
if $\alpha_s\ln[1/\x]$ turns out not to be small for $\x\ll1$.  In the
leading-log approximation the resummation program with respect to such
logarithmic factors has been developed, which brings about running in
x for gluonic properties such as the gluon distribution.  This is
analogous to the famous $Q^2$ (momentum transfer) running of gauge
coupling constant after the summation over leading
$\ln[Q^2/\Lambda^2_{\text{QCD}}]$ terms is taken in perturbative QCD.\
\ Insofar as only the perturbatively calculable part is concerned, it
has been our common knowledge that such running in $\x$, or the
small-$\x$ evolution, in the leading-log approximation obeys the
Balitsky-Fadin-Kuraev-Lipatov (BFKL) equation
\cite{Lipatov:1976zz,Kuraev:1976ge,Kuraev:1977fs,Balitsky:1978ic}.

     The BFKL equation lacks the saturation effects
\cite{Gribov:1984tu,Mueller:1988xy,McLerran:1993ni}, however, that
would be important physics when the parton (gluon) distribution is
highly dense in the small-$\x$ region.  Since the BFKL dynamics
encompasses the evolution in $\x$ but not in $Q^2$, the transverse
area is fixed and the number of partons increases in a fixed area as
$\x$ goes smaller.  Partons eventually overlap each other, indicating
that non-linearity of QCD should be properly taken into account at a
certain saturation scale $Q_s^2(\x)$, and the parton number density
must be saturated via recombination processes.  The modern picture of
this saturation mechanism is more suitably described by coherent gluon
fields $\A_a^\mu$ representing small-$\x$ wee partons
\cite{McLerran:1993ni,McLerran:1993ka,McLerran:1994vd}.  The
larger-$\x$ partons are integrated out to lead to the gluon source
$\rho^a$, from which the coherent fields $\A_a^\mu$ are created
corresponding to smaller-$\x$ degrees of freedom.  [This is a similar
concept to the Weizs\"{a}cker-Williams approximation.]  When
$\A_a^\mu$ becomes as large as $\sim1/g$, non-linearity is crucial for
the saturation.  In the first approximation the source is, originally
in literatures, just regarded as the color charge density carried by
valence quarks inside the target hadron, supposing that the target is
a large nucleus.  Since quarks in different nucleons are uncorrelated,
the color charge density can be considered as distributing randomly.
The random distribution of strong gluon fields $\A_a^\mu$ is commonly
referred to as the color glass condensate \cite{review}.

     One can compute the small-$\x$ evolution, that is actually the
renormalization group equation with the rapidity variable, from $\x$
down to $\x-d\x$ in the presence of the color glass condensate by
assuming that one has such an effective theory of the color glass
condensate at a certain $\x$ and one can retain the same framework at
$\x-d\x$ after integrating out any degrees of freedom within the slice
$d\x=dp^+/P^+$.  The equation obtained in this way is to be considered
as a natural extension of the BFKL equation with the saturation
effect of dense gluons incorporated.  There is an intrinsic energy
scale $Q_s^2(\x)$ which is large, so that the perturbative resummation
becomes infraredly stable \cite{Golec-Biernat:2001if}.  The
theoretical framework has been well organized and the central result
is now widely known as the
Jalilian-Marian-Iancu-McLerran-Weigert-Leonidov-Kovner (JIMWLK)
equation \cite{review,Jalilian-Marian:1997jx,Jalilian-Marian:1997gr,Jalilian-Marian:1997dw,Kovner:2000pt,Iancu:2000hn,Iancu:2001ad,Ferreiro:2001qy}.

     The purpose of this paper is to present the derivation of the
JIMWLK equation in a way different from the conventional arithmetic
steps.  The reasons why we shall attempt to rederive the
well-established equation here are twofold:  The author has recently
proposed a representation for the gauge invariant source terms
\cite{Fukushima:2005kk} which is equivalent with the form as used in
literatures conventionally
\cite{Jalilian-Marian:1997jx,Jalilian-Marian:1997gr,Jalilian-Marian:1997dw,Kovner:2000pt,Iancu:2000hn,Iancu:2001ad,Ferreiro:2001qy},
but is rather suitable for treating the gauge invariance and
non-commutative nature among color charge operators transparently.  It
would be a first-step check to confirm explicitly that such a
representation proposed by the author really leads to the correct
JIMWLK equation in the dense regime, which is necessary as a
preparation for further applications.  Another issue, that could
provide a hint to approach the dilute regime where non-commutativity
among color charge density operators is important, is that in our
formalism we can clearly see under what approximation we have
disregarded non-commutativity in the dense regime.  The JIMWLK
problem, i.e., the derivation of the evolution equation in the dense
regime, has already been well understood itself and our final results
shall not add any new materials to that.  Nevertheless, looking at the
derivation of the JIMWLK equation from another point of view as
addressed in this paper should be useful to understand the JIMWLK
dynamics more deeply and would give us a clue on the treatment beyond
the approximations utilized in the dense regime.  In short, the
ideology of this paper is; the derivation is new, while the derived
equation is already known.

     In Sec.~\ref{sec:cgc} we will quickly look over the philosophy
of the color glass condensate and will articulate our final goal in
advance, namely, the JIMWLK equation.  We will present the central
part of this paper in Sec.~\ref{sec:derivation} where we will explain
a step-by-step derivation of the JIMWLK equation along the formalism
involving quantum color charge density.   Our summary and conclusions
are in Sec.~\ref{sec:conclusion}.  Some discussions are followed in
App.~\ref{sec:conventional} to account for connections between our
derivation and the conventional one for the sake of more clarity.


\section{color glass condensate}
\label{sec:cgc}

     We shall reiterate the physical situation we are considering
here, though we have partially explained it.  A nucleus (target) is
moving fast toward the positive $z$ direction.  In the light-cone
coordinate the target is traveling along the temporal
$x^+=(t+z)/\sqrt{2}$ direction while sitting on
$x^-=(t-z)/\sqrt{2}\simeq0$. In the momentum space language it bears a
large longitudinal momentum $p^+=(p_0+p_z)/\sqrt{2}\sim\infty$.  In
the formulation of the color glass condensate the essential assumption
is that there is a separation scale $\Lambda$ for $p^+$;  all the fast
partons with $|p^+|>\Lambda$ are integrated out to result in the
classical color source $\brho^a_\Lambda(\vec{x})$ which does not
depend on $x^+$ (i.e.\ static) because of the time dilatation relative
to slower partons. The color orientation is assumed to distribute
randomly at each spatial point $\vec{x}=(x^-,\boldsymbol{x}_\perp)$
with a weight $\W_\Lambda[\brho^a_\Lambda]$.  Under these assumptions
the QCD generating functional is modeled as
\begin{equation}
 Z=\int\!\D\brho_\Lambda \W_\Lambda[\brho_\Lambda]\int^\Lambda\!\!
  \D A\,\delta[A^+]\,\exp\Bigl\{i\Sym[A]+iS_W[A^-,\brho_\Lambda]\Bigr\}.
\label{eq:generating}
\end{equation}
It has been shown in \cite{Fukushima:2005kk} that the distribution
$\brho_\Lambda^a(\vec{x})$ is \textit{discrete} at each spatial point
at the microscopic level.  In this paper, however, we shall work only
in the dense regime so that we can approximate
$\brho_\Lambda^a(\vec{x})$ as a continuous function as treated
conventionally.  Here we have employed the light-cone gauge ($A^+=0$)
that enables us to cut-off the longitudinal momentum in a way free
from a mixture of momenta carried by $A^+$ through the covariant
derivative $\partial^+-igA^+$.  It is quite important to
note that $\brho_\Lambda^a(\vec{x})$ has a support in the longitudinal
direction as $0\lesssim x^-\lesssim 1/\Lambda$ as a result of the
integration over fast gluons with an appropriate boundary condition
imposed.  This longitudinal structure is first recognized in
\cite{Iancu:2000hn,Iancu:2001ad,Ferreiro:2001qy} and can be also
embedded in the weight functional $\W_\Lambda[\brho_\Lambda]$ in a way
leading to $\brho_\Lambda^a(\vec{x})=0$ for $x^->1/\Lambda$
\cite{Blaizot:2002xy}.  The JIMWLK equation is, as we will see later,
consistent with the support structure as it should be;  along with the
quantum evolution in $\W_\Lambda[\brho_\Lambda]$ the edge of the
support diffuses to larger $x^-$.  In our notation we do not have to
attach the scale to the source specifically once
$\W_\Lambda[\brho_\Lambda]$ accommodates the support structure , but
we will stick to denoting $\brho_\Lambda^a(\vec{x})$ in this paper to
remind us of the longitudinal structure.

     The support in $x^-$ is intuitively to be interpreted as a
coincidence of the momentum rapidity and the space-time rapidity of
fast partons as argued in \cite{Iancu:2000hn}.  That is, the rapidity
variable,
\begin{equation}
 \tau = \ln\biggl[\frac{1}{\x}\biggr],
\label{eq:rapidity}
\end{equation}
is defined as a momentum rapidity difference between the target and
the parton.  If the particle moves along the longitudinal direction
and its velocity is not changed, which is actually the case in the
eikonal approximation, the momentum rapidity is identical to the
space-time rapidity.  Therefore, if the target is at the position of
$x^-_0$, then the parton with the rapidity difference $\tau$ is
located at $x^-_\tau=x^-_0 e^\tau$.  Since $\brho_\Lambda^a(\vec{x})$
has an origin from fast gluons with $\tau<\ln[P^+/\Lambda]$ in the
color glass condensate picture, its distribution is within
$x^-_0\simeq0\lesssim x^-\lesssim1/\Lambda\simeq x^-_\tau$.  Hereafter
we will often make use of $\tau$ to mean a reference rapidity
$\tau=\ln[P^+/\Lambda]$ instead of using the separation scale
$\Lambda$ like $\W_\tau[\brho_\tau]$ signifying
$\W_\Lambda[\brho_\Lambda]$ equivalently.

     In this paper we expressly call $\brho_\Lambda^a(\vec{x})$ the
\textit{classical} color charge density to distinguish it from the
\textit{quantum} one that fulfills the covariant conservation and the
canonical commutation relations at the operator level.  [We will
explain what this means exactly in the next section.]  In short
$\brho_\Lambda^a(\vec{x})$ turns out to be not a source for quantum
fields but a distribution of fast partons, which is a well-defined
concept at the classical level.  The recognition of
$\brho_\Lambda^a(\vec{x})$ as a classical quantity characterizing the
particle distribution has been discussed already in
\cite{Iancu:2000hn,Iancu:2001ad,Ferreiro:2001qy} where our
$\brho_\Lambda^a(\vec{x})$ is denoted by $\nu$ in order to distinguish
it from the genuine source for quantum fluctuations.  Its significance
has also been emphasized recently in \cite{Fukushima:2005kk}.

     The source action $S_W[A^-,\brho_\Lambda]$ in
(\ref{eq:generating}) is required to be gauge invariant and to be
close to the simple eikonal coupling $\sim\tr\{\brho A^-\}$ not to
affect the equations of motion.  In the most conventional formulation
the source terms have been anticipated as
\begin{equation}
 iS_W[A^-,\brho_\Lambda]=-\frac{1}{gN_c}\int\!d^3x\,\tr\bigl\{
  \brho_\Lambda(\vec{x})W[A^-](\vec{x})\bigr\}.
\label{eq:source_w}
\end{equation}
with the Wilson line along the light-cone temporal direction,
\begin{equation}
 W[A^-](\vec{x})=\mathcal{P}\exp\biggl[\,ig\int_{-\infty}^\infty\!
  dx^+ A^-_a(x^+,\vec{x})\,T^a \biggr],
\label{eq:W}
\end{equation}
defined in the color adjoint representation.  Here the source matrix
is defined in the adjoint basis;
$\brho_\Lambda=\brho_\Lambda^a T^a$.  As usual $\mathcal{P}$ stands
for time (i.e.\ $x^+$) ordering.  The choice of
$S_W[A^-,\brho_\Lambda]$, however, 
is not unique and there are as many variants as one likes that satisfy
gauge invariance and lead to the same equations of motion.  Recently
it has been revealed in \cite{Fukushima:2005kk} that a careful
treatment of the fast gluon integration gives rise to a logarithmic
form,
\begin{equation}
 iS_W[A^-,\brho_\Lambda]=-\frac{1}{gN_c}\int\!d^3x\,\tr\bigl\{
  \brho_\Lambda(\vec{x})\ln W[A^-](\vec{x})\bigr\},
\label{eq:source_log}
\end{equation}
which was proposed first in \cite{Jalilian-Marian:2000ad} motivated by
Wong's equation.  We shall adopt the latter (\ref{eq:source_log})
rather than the conventional choice (\ref{eq:source_w}) throughout
this paper when the explicit form is necessary, though this choice
should be irrelevant to the final results at least within the scope of
the JIMWLK problem.

     Now let us see what follows from the generating functional
(\ref{eq:generating}).  The classical picture of the color glass
condensate results from the tree-level approximation of
(\ref{eq:generating}).  The expectation value of an operator of gauge
fields, $\langle\mathcal{O}[A]\rangle_\Lambda$, is evaluated via
$\langle\langle\mathcal{O}[A]\rangle\rangle_{\brho_\Lambda}$ defined
as
\begin{equation}
 \langle\langle\mathcal{O}[A]\rangle\rangle_{\brho_\Lambda}
  = \frac{1}{z_\Lambda}
  \int^\Lambda\!\D A\,\delta[A^+]\,\mathcal{O}[A]\,
  e^{i\Sym[A]+iS_W[A^-,\brho_\Lambda]}
\end{equation}
with the generating functional
$z_\Lambda[\brho_\Lambda]=\int^\Lambda\!\D A\,\delta[A^+]\,e^{i\Sym+iS_W}$
for states with a given $\brho_\Lambda$.  Then one can estimate
$\langle\mathcal{O}[A]\rangle_\Lambda$ as an average over the
$\brho_\Lambda$-distribution and make a classical approximation;
\begin{equation}
 \langle\mathcal{O}[A]\rangle_\Lambda =\int\!D\brho_\Lambda\,
  \W_\Lambda[\brho_\Lambda]\,
  \langle\langle\mathcal{O}[A]\rangle\rangle_{\brho_\Lambda}
 \simeq \int\!\D\brho_\Lambda\,\W_\Lambda[\brho_\Lambda]\,
  \mathcal{O}\bigl[\A[\brho_\Lambda]\bigr],
\label{eq:expectation}
\end{equation}
using the solution of the Yang-Mills equations of motion
$\A^\mu_a(x)$ as a function of the source,
\begin{equation}
 \frac{\delta\Sym[A]}{\dA^\mu_a(x)}\biggl|_{\A}
  +\frac{\delta S_W[A^-,\brho_\Lambda]}{\dA^\mu_a(x)}\biggl|_{\A}=0.
\end{equation}
This is in fact a stationary-point approximation and expected to be
legitimate when $\W_\Lambda[\brho_\Lambda]$ allows for predominant
contributions from large $\brho_\Lambda\sim 1/g$ and thus large
$\A^\mu_a[\brho_\Lambda]$ in the saturation regime.  Here it is
understood that the weight functional is properly normalized;
$\int\!\D\brho_\Lambda\,\W_\Lambda[\brho_\Lambda]=1$. Since
$\brho_\Lambda$ is just an integration variable in
(\ref{eq:expectation}), the $\Lambda$-dependence in the functional
form of $\W_\Lambda$ solely governs the evolution of the quantity
$\langle\mathcal{O}[A]\rangle_\Lambda$ as $\Lambda$ changes.  This
evolution is what the JIMWLK equation describes.

     The explicit classical solution to the QCD equations of motion,
$D_\nu F^{\nu\mu}-\delta^{\mu+}\brho_\Lambda=0$ (with an ansatz
$\A^-=0$ which makes the source term simple), is given in great
details in literatures \cite{review}.  The covariant gauge
($\partial_\mu \widetilde{A}^\mu=0$ where we follow the convention
that covariant gauge quantities are specifically denoted with a tilde)
has a one-component solution which is as simple as
\begin{equation}
 \widetilde{\A}^\mu[\brho_\Lambda] = \delta^{\mu+}
  \alpha^a_\Lambda(\vec{x})\, T^a,
\end{equation}
where $\alpha^a_\Lambda(\vec{x})$ is a solution of the Poisson
equation in the two-dimensional transverse plane, that is,
\begin{equation}
 \alpha^a_\Lambda(\vec{x})=-\int\!d^2\yt\,\langle\xt|\frac{1}
  {\boldsymbol{\partial}_\perp^2}|\yt\rangle\,
  \brho^a_\Lambda(x^-,\yt)
 =\int\!\frac{d^2\yt}{4\pi} \ln\frac{1}{|\xt-\yt|^2 \mu^2}\,
  \brho^a_\Lambda(x^-,\yt).
\label{eq:poisson}
\end{equation}
Here $\mu^2$ is an infrared regulator that eventually cancels in the
calculation for gauge invariant objects.  The cancellation of $\mu^2$
naturally occurs;  $1/\mu$ gives a scale of long-range force from the
source $\brho_\Lambda$.  The force extends to infinity causing an
infrared singularity at $\mu\to0$.  If an object is color singlet as a
whole and its finer structure inside $\brho_\Lambda^a(\vec{x})$ is not
resolved from a distance, there should be no color force any longer at
distance and thus no infrared singularity.

     The solution in the light-cone gauge ($A^+=0$), which is
necessary in the calculation of quantum fluctuations, is available by
the gauge rotation
$\widetilde{\A}^\mu\to V(\widetilde{\A}^\mu-\partial^\mu/ig)V^\dagger$
eliminating $\widetilde{\A}^+$, that results in
\begin{equation}
 \A^\mu_a[\brho_\Lambda] = -\delta^{\mu i}\,\frac{1}{ig}\,
  V_{x^-,-\infty}(\xt) \, \partial^i \,V_{x^-,-\infty}^\dagger(\xt)
\label{eq:lc_sol}
\end{equation}
with the gauge rotation matrix
\begin{equation}
 V_{x^-,-\infty}^\dagger(\xt)=\mathcal{P}\exp\biggl[\,ig
  \int_{-\infty}^{x^-}\!\!dz^-\alpha^a_\Lambda(z^-,\xt)\, T^a\biggr].
\label{eq:V}
\end{equation}
Here $\mathcal{P}$ represents the ordering in $x^-$ unlike
$x^+$-ordering in (\ref{eq:W}).  In this paper, in order to avoid
tangled notations like $\widetilde{\brho}_\Lambda$, we would use only
$\brho_\Lambda$ with a remark about the gauge choice if necessary.  In
the above discussion $\brho_\Lambda$ is all in the covariant gauge,
while the light-cone gauge solution (\ref{eq:lc_sol}) satisfies the
equations of motion for the source in the light-cone gauge, namely,
$V_{x^-,-\infty}\brho_\Lambda V^\dagger_{x^-,-\infty}$ written in
terms of the covariant gauge source.

     From these relations we can choose to use
$\alpha^a_\Lambda(\vec{x})$ to express a function of
$\brho^a_\Lambda(\vec{x})$.  Since $\brho^a_\Lambda(\vec{x})$ has the
support over $0\lesssim x^-\lesssim1/\Lambda$, so does
$\alpha^a_\Lambda(\vec{x})$.  One might seem perplexed at this at
first glance;  since the $x^-$-structure is such localized,
$\alpha^a_\Lambda(\vec{x})$ carries a large longitudinal momentum
$|p^+|>\Lambda$ like fast partons generating $\brho_\Lambda(\vec{x})$.
At the same time the classical fields should describe wee partons with
$|p^+|<\Lambda$.  This is not a contradiction, however.  The point is
that the scale separation in $p^+$, as we have mentioned, makes sense
not in the covariant gauge but in the light-cone gauge only.  Actually
$\A^i[\brho_\Lambda]$ in the light-cone gauge is delocalized and takes
a nonvanishing value at $x^-\gtrsim1/\Lambda$.  [It should be noted
that the retarded boundary condition is implicit in the choice of
(\ref{eq:V}) so that only positive $x^-$ is taken.]  Recalling a
coincidence between the momentum and space-time rapidities, the
light-cone gauge solution $\A^i[\brho_\Lambda]$ distributing over
$x^-\gtrsim1/\Lambda$ surely corresponds to slow partons.

     We are now ready to write the JIMWLK equation down in a
well-known form in terms of $\alpha^a_\tau(\vec{x})$ using the
rapidity variable $\tau$ defined in (\ref{eq:rapidity});
\begin{equation}
 \frac{\partial}{\partial\tau} \W_\tau[\alpha_\tau] = \frac{1}{2}
  \int\!d^2\xt d^2\yt\, \frac{\delta}
  {\delta\alpha^a_\tau(x^-_\tau,\xt)}\eta^{ab}(\xt,\yt)\frac{\delta}
  {\delta\alpha^b_\tau(x^-_\tau,\yt)} \W_\tau[\alpha_\tau],
\label{eq:JIMWLK}
\end{equation}
where the derivatives act only on the edge of the support
$x^-=x^-_\tau$ and
\begin{equation}
 \eta^{ab}(\xt,\yt)=\frac{1}{\pi}\int\!\frac{d^2\zt}{(2\pi)^2}\,
  \K(\xt,\yt,\zt)\Bigl[1-V^\dagger(\zt)V(\yt)-V^\dagger(\xt)V(\zt)
  +V^\dagger(\xt)V(\yt)\Bigr]^{ab}
\label{eq:eta}
\end{equation}
with the transverse kernel,
\begin{equation}
 \K(\xt,\yt,\zt)=(2\pi)^2 \langle\zt|\frac{\partt}{\partt^2}|\xt\rangle
  \cdot \langle\zt|\frac{\partt}{\partt^2}|\yt\rangle
 =\frac{(\xt-\zt)\cdot(\yt-\zt)}{(\xt-\zt)^2(\yt-\zt)^2}.
\end{equation}
Here $V^\dagger(\xt)=V^\dagger_{x^-_\tau,-\infty}(\xt)\simeq
V^\dagger_{+\infty,-\infty}(\xt)$ with the definition (\ref{eq:V}).

     The derivation of the JIMWLK equation (\ref{eq:JIMWLK}) is our
goal pursued in this paper.  Usually one derives the JIMWLK equation
from the fact that the generating functional (\ref{eq:generating}) is
independent of the separation scale $\Lambda$ as we will explain in
some details in App.~\ref{sec:conventional}.  We shall here take a
different starting point:  The important observation pointed out in
\cite{Fukushima:2005kk} is that the generating functional is written
equivalently in a different form with an auxiliary variable $\rho$ as
\begin{equation}
 Z=\int\!\D\rho\,\N_\Lambda[\rho]\int^\Lambda\!\!\D A\,\delta[A^+]\,
  \exp\biggl[\,i\Sym[A]-i\int\!d^4x\,\rho^a(x) A^-_a(x)\biggr]
\label{eq:generating_n}
\end{equation}
with the density of states $\N_\Lambda[\rho]$ defined as
\begin{equation}
 \N_\Lambda[\rho]=\int\!\D\brho_\Lambda \W_\Lambda[\brho_\Lambda]\,
  \exp\biggl[-\frac{1}{gN_c}\int\!d^3x\,\tr\bigl\{
  \brho_\Lambda(\vec{x}) \ln W[-i\delta/\delta\rho](\vec{x})\bigr\}
  \biggr]\delta[\rho].
\label{eq:density}
\end{equation}
One can immediately prove that (\ref{eq:generating_n}) returns exactly
to (\ref{eq:generating}) after the $\rho$-integration.  Because
(\ref{eq:generating_n}) is mathematically equivalent with
(\ref{eq:generating}), one should be able to derive the same evolution
equation from the $\Lambda$-independence of $Z$ as written in
(\ref{eq:generating_n}).  We are attempting to make sure this
expectation in what follows.  Let us see as an instant check how the
equivalent classical picture of the color glass condensate arises.
This is simply concluded from a claim that $\rho^a(x)$ is static,
which corresponds to an approximation that non-commutativity is
disregarded as we will argue later (see discussions around
(\ref{eq:exp_lnW})).  There is then no discrimination between
$\brho_\Lambda^a(\vec{x})$ and $\rho^a(x)$ and moreover
$\N_\Lambda[\rho]$ is reduced to $\W_\Lambda[\brho_\Lambda]$ and two
representations become trivially identical.  The quantum evolution is,
however, not so trivial.  The derivation of the JIMWLK equation based
on (\ref{eq:generating_n}) is the main issue of this paper and
elucidated in Sec.~\ref{sec:derivation}.


\section{derivation with quantum color charge density}
\label{sec:derivation}

     We will derive the JIMWLK equation (\ref{eq:JIMWLK}) with our
emphasis on the quantum nature of color charge operators.  We would
remark here again that the \textit{derivation} presented in this
section is our central achievement of this paper.  First in
Sec.~\ref{sec:quantum} we will explain about non-commutativity among
quantum color charge density from the field theoretical point of view,
and then we will make proceed in the concrete calculus in
Sec.~\ref{sec:renormalization}.


\subsection{Quantum color charge density}
\label{sec:quantum}

     Before addressing the computational steps, let us clarify the
meaning of \textit{quantum} color charge density and explain why we
are interested in that.  In the canonical quantization, in gauge
theories, the color charge density operators obey the commutation
relation at equal time,
\begin{equation}
 \bigl[\hat{\rho}^a(x^+,\vec{x}),\hat{\rho}^b(x^+,\vec{y})\bigr] =
  -igf^{abc}\hat{\rho}^c(x^+,\vec{x})\;\delta^{(3)}(\vec{x}-\vec{y}),
\label{eq:commutation}
\end{equation}
which one can easily confirm by using the canonical quantization
conditions.  The commutation relation (\ref{eq:commutation}) holds for
the source of any (fermionic or bosonic) matter field.  In the
operator formalism of quantization, (\ref{eq:commutation}) is
necessary for consistency between Gauss' law and gauge invariance,
that is, the secondary constraint in the nonabelian gauge theory
(i.e.\ Gauss' law) is first-class if we use Dirac's nomenclature for
quantization of singular systems~\cite{Dirac}.  If the theory is
formulated in the functional integral formalism,
(\ref{eq:commutation}) is expressed by the expectation values in terms
of classical fields.  In the functional integral formalism, by
construction, the expectation value is taken in a time-ordered way and
(\ref{eq:commutation}) is reexpressed as
\begin{equation}
 \bigl\langle\mathcal{O}[\rho]\bigl\{\rho^a(x^+,\vec{x})
  \rho^b(x^+\!-\eta,\vec{y})-\rho^b(x^+,\vec{y})\rho^a(x^+\!
  -\eta,\vec{x})\bigr\}\bigr\rangle \Bigl|_{\eta\to0^+}
 = -igf^{abc}\bigl\langle\mathcal{O}[\rho]\,\rho^c(x^+,\vec{x})
  \bigr\rangle\; \delta^{(3)}(\vec{x}-\vec{y})
\label{eq:commut_integ}
\end{equation}
for arbitrary functionals $\mathcal{O}[\rho]$.

     In the JIMWLK problem relevant in the dense regime where the
color charge density is as large as $\sim1/g$, we can ignore the
right-hand side of (\ref{eq:commut_integ}).  Then, non-commutativity
does not play any significant role in the JIMWLK evolution equation
(apart from generating non-local vertices corresponding to the
expansion of the source terms \cite{Hatta:2005wp}).  It is a
challenging problem to transcend the JIMWLK equation towards the
dilute regime where (\ref{eq:commutation}) or (\ref{eq:commut_integ})
must be taken into account.  This attempt has been in part successful
so far;  the expansion in terms of the charge density derivative has
been improved up to the forth order \cite{Mueller:2005ut} and now the
general Hamiltonian for all orders has been known
\cite{Kovner:2005nq,Hatta:2005rn}, while the complete quantization
procedure has not been quite clear yet.  In the dilute regime the
quantum evolution has also been studied in the color dipole
picture \cite{Mueller:1993rr} which can be connected to the color
glass formalism \cite{Iancu:2003uh}.

     In the operator formalism the Heisenberg equation with
(\ref{eq:commutation}) describes the evolution, given the Hamiltonian
is known as a function of the charge density operators.  Since the
color glass formulation, as we have seen in Sec.~\ref{sec:cgc}, is
naturally understood in the form of the functional integral, it is
desired to formulate the evolution equation consistent with
(\ref{eq:commut_integ}) within the functional integral description.
Recalling that the color charge operators can be regarded as
\textit{spin} operators aligned in color group space, one can
anticipate a functional integral representation from knowledge on the
spin system.  That is, the action for the spin system is known to
consist of the dynamical and geometrical parts.  In the language of
quantum mechanics, the action is the integral of the Lagrangian which
is given as $L(q)=p\dot{q}-H(p)$ with a variable $q$ and its canonical
conjugate $p$.  The dynamical and geometrical parts correspond to the
integral of $H(p)$ and $p\dot{q}$ respectively.  In fact the
geometrical part has an interpretation as the topological phase which
is often referred to as Berry's phase.  Therefore, some attempts to
augment the formalism with Berry's phase in color space have been
proposed in order to handle (\ref{eq:commut_integ}) properly
\cite{Kovner:2005nq,Hatta:2005wp}.

     Recently, in \cite{Fukushima:2005kk}, a different way to deal
with non-commutativity (\ref{eq:commut_integ}) has been proposed in
the functional integral formulation.  The advantage of that proposed
method as compared with the representation with Berry's phase is that
the necessary modification to the conventional formulation is
minimal.  There is no need to \textit{add} Berry's phase into the
formulation.  The point is that $\rho^a(x)$ appearing in
(\ref{eq:generating_n}) rather than $\brho_\Lambda^a(\vec{x})$
fulfills (\ref{eq:commut_integ}) and can be regarded as quantum color
charge density.  That is simply because the color charge density
operator corresponds to $i\delta/\delta A^-_a(x)$ acting on the source
action $\exp[iS_W[A^-,\rho]]$.  If we choose the source action as in
(\ref{eq:generating_n}), then $i\delta/\delta A^-_a(x)$ is replaced by
$\rho^a(x)$ coming from the eikonal coupling.  It is obvious at this
point that non-commutativity is lost if the $x^+$-dependence of
$\rho^a(x)$ is dropped.  In the next subsection we will see how this
$x^+$-dependence is eventually abandoned in deriving the JIMWLK
equation.  This means, in other words, that one would be able to
surpass the JIMWLK equation if one keeps the full $x^+$-dependence,
which is beyond the scope of this present paper.


\subsection{Renormalization group equation}
\label{sec:renormalization}

     We shall elaborate the renormalization group equation here from
the fact that the density of states $\N_\Lambda[\rho]$ evolves so as
to keep the whole generating functional $Z$ independent of the
separation scale $\Lambda$.  That is, $Z$ given in
(\ref{eq:generating_n}) can be written at a different separation scale
$b\Lambda$ for arbitrary $b$.  In this paper $b$ is assumed to be
small as usual, i.e.\ $b\ll1$ and we are working up to the leading
order of $\alpha_s\ln[1/b]\sim\alpha_s d\tau$ and will finally take
the limit of $b\to1$ to acquire the differential equation in $\tau$.
Then one will reach $\N_{b\Lambda}[\rho]$ expressed in terms of
$\N_{\Lambda}[\rho]$ accompanied by contributions from the integration
over degrees of freedom inside the scale from $\Lambda$ to $b\Lambda$.
Such a relation is nothing but the renormalization group equation for
$\N_\Lambda[\rho]$ that flows according to the scale $\Lambda$ or the
rapidity $\tau$.

     Let us briefly summarize in advance each step for the calculation
procedure for the sake of guiding what will transpire below.
\vspace{2mm}

     (Step-1)\ \ We will first start with the generating
functional $Z$ written at the separation scale $\Lambda$.  We will
decompose the gauge fields into the slow and fast parts according to
the longitudinal momentum $p^+$, and we will expand the Yang-Mills
action up to the quadratic order in semi-fast gluons $a^\mu_a$ which
we will define there.
\vspace{1mm}

     (Step-2)\ \ We will perform the Gaussian integration with respect
to $a^\mu_a$ and will reach an expression of the generating functional
$Z$ with the gauge field fluctuation below the scale $b\Lambda$.
\vspace{1mm}

     (Step-3)\ \ We will equate the counterpart of the density of
states with $\N_{b\Lambda}[\rho]$ after the Gaussian integration.  In
the limit of $b\to1$ we will acquire the differential equation, that
is, the renormalization group equation.
\vspace{1mm}

     (Step-4)\ \ We will rewrite the renormalization group equation in
terms of $\W_\Lambda[\brho_\Lambda]$.
\vspace{2mm}

     In the rest of this subsection we shall closely explain each step
in order.  In App.~\ref{sec:conventional}, for comparison, we will
sketch the conventional derivation of the JIMWLK equation directly for
$\W_\Lambda[\brho_\Lambda]$ in accord with (Step-1), (Step-2), and
(Step-3) above.
\vspace{5mm}


     (Step-1)\ \ In order to eliminate degrees of freedom within the
momentum shell to be integrated, we shall decompose the gauge fields
into the semi-fast gluons $a^\mu_a$ whose longitudinal momenta lie in
the strip, $b\Lambda<|p^+|<\Lambda$, the classical solution,
$\A^\mu_a[\brho_{b\Lambda}]$ of (\ref{eq:lc_sol}), and the rest
$\dA^\mu_a$ with $|p^+|<b\Lambda$, that is,
\begin{equation}
 A^\mu_a \to a^\mu_a + \delta^{\mu i}\A^i_a[\brho_{b\Lambda}]
  + \delta^{\mu -}\dA^-_a \,.
\label{eq:decompose}
\end{equation}
We assume that the transverse fluctuation around $\A^i_a$ is small
enough to be negligible.  Actually one can show that the only current
(defined as the conjugate to $\dA^\mu_a$) correlators to be retained
in the leading-log approximation are correlators of the charge density
(conjugate to $\dA^-_a$) \cite{Iancu:2000hn}.  Therefore the specific
decomposition (\ref{eq:decompose}) is validated under the present
approximation.  It should be noted that from the argument below
(\ref{eq:V}) we know that the classical solution appears as a result
of the stationary point approximation in evaluating the generating
functional, and thus the classical part in the above decomposition
(\ref{eq:decompose}) should be that for $\brho_{b\Lambda}$ (not for
$\brho_\Lambda$).  This is because $a^\mu_a$ is integrated out and
$\A^i_a[\brho_{b\Lambda}]$ should correspond to the stationary point
of fields which have $|p^+|<b\Lambda$.

     Then, it is straightforward to expand the Yang-Mills action in
the generating functional at the scale $\Lambda$ in terms of $a^\mu_a$
up to the quadratic order.
\begin{equation}
 \begin{split}
 & Z \simeq\int\!\D\rho\,\N_\Lambda[\rho]\int^{b\Lambda}\!\!\D\dA\,
  \delta[\dA^+]\,\exp\biggl[\,i\Sym[\A[\brho_{b\Lambda}]+\dA]-i\int\!
  d^4x\,\rho^a(x) \dA^-_a(x)\biggr]\\
 &\times \int_{b\Lambda}^\Lambda\!\!\D a\,\delta[a^+]\,\exp\biggl[\,
  i\int\!d^4x\,\frac{\delta\Sym}{\dA^\mu_a(x)}\, a^\mu_a(x) +\frac{i}{2}
  \int\!d^4x\,d^4y\,\frac{\delta^2\Sym}{\dA^\mu_a(x) \dA^\nu_b(y)}\,
  a^\mu_a(x)a^\nu_b(y) -i\int\!d^4x\,\rho^a(x)a^-_a(x)\biggr],
 \end{split}
\label{eq:expansion}
\end{equation}
where the derivatives of the Yang-Mills action are taken at
$\A^i_a[\brho_{b\Lambda}]+\dA^-_a$.
\vspace{3mm}


     (Step-2)\ \ Since (\ref{eq:expansion}) is quadratic in $a^\mu_a$
one can easily perform the Gaussian integration to have
\begin{align}
 &\int_{b\Lambda}^\Lambda\!\D a\,\delta[a^+]\,\exp\biggl[-\frac{1}{2}
  \int\!d^4x\,d^4y\,\biggl\{a^\mu_a(x)-i\int\!d^4z\biggl(
  \frac{\delta \Sym}{\dA^\lambda_c(z)} -\delta^{\lambda-}\rho^c(z)
  \biggr) G^{\lambda\mu}_{0ca}(z,x)\biggr\}\, G_{0\mu\nu}^{-1ab}(x,y)
  \notag\\
 &\qquad\times  \biggl\{a^\nu_b(y)-i\int\!d^4z'
  G^{\nu\lambda'}_{0bc'}(y,z') \biggl(\frac{\delta \Sym}
  {\dA^{\lambda'}_{c'}(z')}-\delta^{\lambda'-}\rho^{c'}(z')\biggr)
  \biggr\}  \notag\\
 &\qquad\qquad -\frac{1}{2}\int\!d^4x\,d^4y\,\biggl(\frac{\delta \Sym}
  {\dA^\mu_a(x)}-\delta^{\mu-}\rho^a(x)\biggr) G^{\mu\nu}_{0ab}(x,y)
  \biggl(\frac{\delta\Sym}{\dA^\nu_b(y)}-\delta^{\nu-}\rho^b(y)\biggr)
  \biggr] \notag\\
 =& \det\Bigl[ G^{-1ab}_{0\mu\nu}(x,y) \Bigr]^{-\frac{1}{2}}
  \exp\biggl[-\frac{1}{2}\int\!d^4x\,d^4y\,\biggl(\frac{\delta\Sym}
  {\dA^\mu_a(x)}-\delta^{\mu-}\rho^a(x)\biggr) G^{\mu\nu}_{0ab}(x,y)
  \biggl(\frac{\delta \Sym}{\dA^\nu_b(y)}-\delta^{\nu-}\rho^b(y)
  \biggr) \biggr],
\label{eq:gaussian}
\end{align}
where the propagator inverse for the semi-fast gluon is written as
\begin{equation}
 iG^{-1ab}_{0\mu\nu}(x,y)=\frac{\delta^2\Sym}{\dA^\mu_a(x)\dA^\nu_b(y)}
  =\Bigl\{\bigl[D_\lambda D^\lambda\bigr]^{ab}g_{\mu\nu}-\bigl[
  D_\mu D_\nu\bigr]^{ab}+2gf^{acb}F_{\mu\nu}^c\Bigr\}\,
  \delta^{(4)}(x-y)
\end{equation}
with
\begin{equation}
 D^-=\partial^--ig\dA^-,\quad D^i=\partial^i-ig\A^i,\qquad
 F_{+i}=-D_i\dA^-,\quad F_{-i}=\partial^+\A_i,\quad F_{ij}=0
\label{eq:d_f}
\end{equation}
in the light-cone gauge.  We note that all quantities listed in
(\ref{eq:d_f}) are matrices in the color adjoint basis.

     In the rest of this step let us evaluate the determinant in the
above expression (\ref{eq:gaussian}), namely, the determinant of the
propagator inverse which is expressed in the ($a^-,a^i$) basis
explicitly as
\begin{equation}
 iG^{-1ab}_{0\mu\nu}(x,y)=\left[ \begin{array}{cc}
-\partial^{+2}\delta^{ab} &
 -\partial^+ D_j^{ab}+2gf^{acb}(\partial^+\A^c_j) \\
-D_i^{ab}\partial^+-2gf^{acb}(\partial^+\A^c_i) &
 -[D_\lambda D^\lambda]^{ab}\delta_{ij}-[D_i D_j]^{ab}
  \end{array} \right] \,,
\end{equation}
from which one can immediately write the determinant of matrix blocks
as
\begin{equation}
 \begin{split}
 &\det\Bigl[G^{-1ab}_{0\mu\nu}(x,y)\Bigr]^{-\frac{1}{2}}
  =\det[-\partial^{+2}\delta^{ab}]^{-\frac{1}{2}}  \\
 &\quad \times \det\Bigl[-[D_\lambda D^\lambda]^{ab}\delta_{ij}
  -[D_i D_j]^{ab}-\bigl\{-D_i^{ad}\partial^+-2gf^{acd}
  (\partial^+\A^c_i)\bigr\}(-\partial^{+2})^{-1}\bigl\{-\partial^+
  D_j^{db}+2gf^{dc'b}(\partial^+\A^{c'}_j)\bigr\}\Bigr]^{-\frac{1}{2}}.
 \end{split}
\end{equation}
Here the first part, $\det[-\partial^{+2}\delta^{ab}]^{-\frac{1}{2}}$,
is just an irrelevant number, and after some algebra the second part
simplifies as
\begin{equation}
 \begin{split}
 &\det\biggl[-[D_\lambda D^\lambda]^{ab}\delta_{ij} -2gf^{acb}
  \biggl\{\frac{\partial_i}{\partial^+}(\partial^+\A^c_j)
  -(\partial^+\A^c_i)\frac{\partial_j}{\partial^+}\biggr\}
  -2g^2 f^{acd}f^{dc'b}\biggl\{\A_i^c\frac{1}{\partial^+}
  (\partial^+\A^{c'}_j)-(\partial^+\A^c_i)\frac{1}{\partial^+}
  \A^{c'}_j\biggr\} \\
 &\qquad\qquad -4g^2f^{acd}f^{dc'b}(\partial^+\A^c_i)
  \frac{1}{\partial^{+2}}(\partial^+\A^{c'}_j)\biggr]^{-\frac{1}{2}}.
 \end{split}
\label{eq:determinant}
\end{equation}
Although this complicated expression looks impracticable to proceed
further, we can drop many terms under the leading-log approximation as
we will explain now. Let us focus on the second term
$\sim(\partial_i/\partial^+)(\partial^+\A_j)$ (the first term in the
first curly brackets) for the moment to take an example.  If one
remembers that the determinant originates from the terms sandwiched
by the semi-fast gluons $a^\mu_a$, one readily understands that the
derivative $\partial_i/\partial^+$ picks up the momenta associated
with the semi-fast gluons, which is proportional to $1/p^+$ with
$b\Lambda<|p^+|<\Lambda$.  The terms suppressed by $1/p^+$ can be
neglected because no logarithmic enhancement $\sim\ln[1/b]$ arises
from those terms.  Because the first term
$-[D_\lambda D^\lambda]^{ab}\delta_{ij}$ contains a term proportional 
to $p^+$, the logarithmic enhancing terms could appear from the part
with no such $1/p^+$ suppression if the determinant is expanded in the
power of $1/p^+$.  Let us next turn to the second curly brackets in
(\ref{eq:determinant}) which contains
$\A_i(1/\partial^+)(\partial^+\A_j)\sim\A_i\A_j$ apart from the
suppressed term $\A_i(\partial^+\A_j)(1/\partial^+)$ that is to be
dropped by the reason we now explained.  This would potentially
produce logarithmic enhancing terms if expanded.  Interestingly
enough, however, there is a complete cancellation by the contribution
from the last term in (\ref{eq:determinant}), i.e.\
$(\partial^+\A_i)(1/\partial^{+2})(\partial^+\A_j)\sim\A_i\A_j$ apart
from the suppressed terms.  Consequently we only have to retain the
first term in (\ref{eq:determinant}) and thus arrive at
\begin{equation}
 \det\Bigl[-[D_\lambda D^\lambda]^{ab}\delta_{ij}
  \Bigr]^{-\frac{1}{2}},
\end{equation}
within the leading-log approximation.  Now we make a further
approximation on this.  By the argument given in \cite{Hatta:2005rn}
the interaction region in which both $\brho_{\Lambda}(\vec{x})$ and
$\dA^-_a(x)$ are nonvanishing is presumably small in the
two-dimensional space spanned by $x^+$ and $x^-$, that means both
$p^+$ and $p^-$ are substantially larger than the transverse momentum
scale.  Then we can ignore $\partial_i\partial_i$ as compared with
$2\partial^+\partial^-$ inside the determinant.  Using
$\partial^+\dA^-\simeq0$ which is valid in the light-cone gauge in the
leading-log approximation ($\dA^-$ has $|p^+|<b\Lambda$ and so
$\partial^+\dA^-$ leads to contributions $\propto b$ which are
neglected) the determinant is thus
\begin{equation}
 \det\Bigl[G^{-1ab}_{0\mu\nu}(x,y)\Bigr]^{-\frac{1}{2}}\simeq
  \det\bigl[2\widetilde{D}^+ \widetilde{D}^-\bigr]^{-1}
  = \det\bigl[\partial^+-ig\alpha\bigr]^{-1}
  \det\bigl[\partial^--ig\dA^-\bigr]^{-1}.
\label{eq:det2}
\end{equation}
Here we once moved to the covariant gauge to avoid the complexity
associated with entanglement between $\partial^i$ which is neglected
and $\A^i$ in the light-cone gauge.  It should be mentioned here that
the determinant of the covariant derivatives as in (\ref{eq:det2}) is
\textit{gauge independent} as it is individually.  Thus we
could rotate $\delta\widetilde{A}^-$ in the covariant gauge back to
$\dA^-$ in the light-cone gauge without touching anything on
$\det[\partial^+-ig\alpha]^{-1}$, namely,
$\det[\partial^+-ig\dA^-]=\det[\partial^+-ig\delta\widetilde{A}^-]$.
In this paper we do not discuss but just discard
$\det[\partial^+-ig\alpha]$ which would not play any role in the
JIMWLK problem but might be important if we consider the evolution of
not only the target but also the projectile.

     The fact that $\det[\partial^--ig\dA^-]$ is gauge independent
implies that this be reduced to an irrelevant number if $\dA^-$ is
eliminated by an appropriate gauge rotation.  In other words, the
determinant is expanded as a sum of all the Wilson \textit{loops}
which turn out to be irrelevant numbers because they sit only on the
one-dimensional $x^+$ space.  If we require the periodic boundary
condition in $x^+$, which could be suggested from the physical
situation of elastic scattering, we can consider the non-trivial
Wilson loop wrapping over $x^+$ whose starting and ending points are
connected by periodicity.  Therefore we can expect that under the
periodic boundary condition $\det[\partial^--ig\dA^-]$ is given as a
function of $W[\dA^-]$ defined in (\ref{eq:W}).

     The explicit evaluation of (\ref{eq:det2}) is possible in the
same way as discussed in \cite{Fukushima:2005kk}, leading to the the
classical color charge density as
\begin{equation}
 \det\bigl[\partial^--ig\dA^-\bigr]^{-1} = \int\!\D\delta\brho_\Lambda\,
  w_\Lambda[\delta\brho_\Lambda]\, \exp\biggl[
  -\frac{1}{gN_c}\int\!d^3x\,\tr\bigl\{\delta\brho_\Lambda(\vec{x})
  \ln W[\dA^-](\vec{x})\bigr\}\biggr].
\label{eq:renorm}
\end{equation}
This is a gauge invariant expression, as we have already noted.  Here
the correction to the classical color charge density
$\delta\brho_\Lambda(\vec{x})$ distributes in
$1/\Lambda<|x^-|<1/b\Lambda$ because the determinant is taken in the
space spanned by the semi-fast gluon which has
$b\Lambda<|p^+|<\Lambda$ by definition.  Usually only the positive
$x^-$ distribution is chosen in accord to the retarded boundary
condition which is considered as physically relevant.  The
combinatorial weight function associated with the spatial
distribution is denoted by $w_\Lambda[\delta\brho_\Lambda]$.  This
part, however, has no logarithmic enhancement $\sim\ln[1/b]$ and is to
be discarded in the leading-log approximation.  It should be mentioned
here that the term containing $\ln W[\dA^-]$ turns out to be a shift
operator acting on logarithmically enhanced terms and thus we should
keep it.
\vspace{3mm}

     (Step-3)\ \  Now that we have accomplished the Gaussian
integration, we can identify the counterpart corresponding to
$\N_{b\Lambda}[\rho]$ as a function of $\N_\Lambda[\rho]$, which leads
to the renormalization group equation for $\N_\Lambda[\rho]$;
\begin{equation}
 \begin{split}
 & \int\!\D\rho\,\N_{b\Lambda}[\rho]\,e^{-i\int\!d^4x\,\rho^a(x)
  \dA^-_a(x)} = \int\!\D\rho\,\N_\Lambda[\rho]\,\exp\bigl[\Delta
  S[\A[\brho_{b\Lambda}]+\dA^-,\rho]\bigr] \\
 &\qquad\qquad\qquad\qquad\times \int\!\D\delta\brho_\Lambda\,
  \exp\biggl[-\frac{1}{gN_c}\int\!d^3x\,\tr\bigl\{
  \delta\brho_\Lambda(\vec{x}) \ln W[i\delta/\delta\rho](\vec{x})
  \bigr\}\biggr]\,e^{-i\int\!d^4x\,\rho^a(x)\dA^-_a(x)},
 \end{split}
\label{eq:evolv_N}
\end{equation}
where $\Delta S$ is the rest of the Gaussian integration
(\ref{eq:gaussian}) given by
\begin{equation}
 \Delta S[\A[\brho_{b\Lambda}]+\dA^-,\rho] =-\frac{1}{2}\biggl(
  \frac{\delta\Sym}{\dA^\mu_a(x)}-\delta^{\mu-}\rho^a(x)\biggr)
  G^{\mu\nu}_{0ab}(x,y) \biggl(\frac{\delta \Sym}{\dA^\nu_b(y)}
  -\delta^{\nu-}\rho^b(y)\biggr).
\label{eq:delS}
\end{equation}

(Step-4)\ \ Then the evolution equation for the weight functional
$\W_\Lambda[\brho_\Lambda]$ follows immediately from
(\ref{eq:evolv_N}) as
\begin{equation}
 \begin{split}
 &\int\!\D\rho\;e^{-i\int\!d^4x\,\rho^a(x)\dA^-_a(x)} \int\!
  \D\brho_{b\Lambda}\,\biggl[\W_{b\Lambda}[\brho_{b\Lambda}]\,e^{
  -\frac{1}{gN_c}\int\!d^3x\,\tr\{\brho_{b\Lambda}(\vec{x})
  \ln W[-i\delta/\delta\rho](\vec{x})\}} \\
 &\qquad -\W_{\Lambda}[\brho_{b\Lambda}]\,e^{-\frac{1}{gN_c}\int\!d^3x\,
  \tr\{\delta\brho_\Lambda(\vec{x})\ln W[-i\delta/\delta\rho](\vec{x})\}}
  \,e^{\Delta S[\A[\brho_{b\Lambda}]+\dA^-,\rho]}\,e^{-\frac{1}{gN_c}
  \int\!d^3x\,\tr\{\brho_\Lambda(\vec{x})\ln W[-i\delta/\delta\rho]
  (\vec{x})\}}\biggr]\delta[\rho] =0.
 \end{split}
\label{eq:evolv_W}
\end{equation}
Here we defined the integration
$\int\!\D\brho_{b\Lambda}=\int\!\D\brho_\Lambda\int\!\D\delta\brho_\Lambda$,
and the weight functional is extended as \cite{Blaizot:2002xy}
\begin{equation}
 \W_\Lambda[\brho_{b\Lambda}] = \W_\Lambda[\brho_\Lambda]
  \prod_{1/\Lambda<x^-<1/b\Lambda} \delta[\drho_\Lambda].
\end{equation}
We leave $\dA^-$ in $\Delta S$ as it is (not replaced by
$-i\delta/\delta\rho$) on purpose to avoid confusion in the
approximation we will make later.

     In the dense regime where $\brho_\Lambda$ takes a large value,
we can make an important approximation here, i.e.\ we can expand the
Wilson line term as
\begin{align}
 & -\frac{1}{gN_c}\int\!d^3x\,\tr\bigl\{\brho_\Lambda(\vec{x})
  \ln W[-i\delta/\delta\rho](\vec{x})\bigr\} \notag\\
 \simeq & -\int\!d^4x\,\brho_\Lambda^a(\vec{x})\frac{\delta}
  {\delta\rho^a(x)} -\frac{g}{2N_c}\int\!d^3x\,(\mathcal{P}-1)\int\!
  dx^+dy^+\, \tr\biggl\{ \brho_\Lambda(\vec{x}) \frac{\delta}
  {\delta\rho(x^+)}\frac{\delta}{\delta\rho(y^+)} \biggr\} +\cdots ,
\label{eq:exp_lnW}
\end{align}
where the higher-order contributions involve more derivatives
$\delta/\delta\rho$, so that they are suppressed when $\brho_\Lambda$
or $\rho$ is large.  [$\rho$ is of the same order as $\brho_\Lambda$
as we will see shortly.]  Here $\mathcal{P}$ stands for time ordering
as in (\ref{eq:W}).  The first term in (\ref{eq:exp_lnW}) placed on
the exponential leads to the shift operator which makes $f[\rho]$ be
$f[\rho-\brho_\Lambda]$.  The second term is vanishing under the
second-order perturbation that the terms up to the quadratic order in
$\dA^-$ or $\delta/\delta\rho^a$ are kept;  the derivatives in the
second term should act on the function of $\rho^a(x)$ arising as a
result of the operation by the first term.  Then the delta functional
shifted by $\brho_\Lambda^a(\vec{x})$ leads to
$\rho^a(x)=\brho_\Lambda^a(\vec{x})$ which is $x^+$ independent.  Once
$\rho^a(x)$ loses the $x^+$ dependence, then only the first term in
the expansion (\ref{eq:exp_lnW}) remains, which signifies that
non-commutativity becomes completely irrelevant.  Therefore we insist
that the expansion (\ref{eq:exp_lnW}) is the key equation to
understand how the non-commutativity is disregarded in the JIMWLK
problem in the high density regime.  In App.~\ref{sec:conventional} we
will comment on the connection of (\ref{eq:exp_lnW}) to an
approximation made in the conventional derivation of the JIMWLK
equation.

     In evaluating (\ref{eq:evolv_W}), keeping only the first term in
the expansions like (\ref{eq:exp_lnW}), one can simplify it as
\begin{equation}
 \int\!\D\rho\;e^{-i\int\!d^4x\,\rho^a(x)\dA^-_a(x)} \int\!\D
  \brho_{b\Lambda}\,\biggl[\W_{b\Lambda}[\brho_{b\Lambda}]
  -\W_\Lambda[\brho_{b\Lambda}]\,e^{\Delta S[\A[\brho_{b\Lambda}]+\dA^-,
  \rho+\drho_\Lambda]}\biggr]\delta[\rho-\brho_\Lambda-\delta
  \brho_\Lambda] =0.
\end{equation}
One can readily complete the $\rho$-integration here, then replace
$\dA^-$ in $\Delta S$ by the derivative in
$\brho_{b\Lambda}=\brho_\Lambda+\drho_\Lambda$, and move the
derivative onto $\W_\Lambda[\brho_{b\Lambda}]$ by the integration by
parts.  Finally we get the evolution equation in the form of
\begin{equation}
 \W_{b\Lambda}[\brho_{b\Lambda}]- e^{\Delta S[\A[\brho_{b\Lambda}]
  -i\delta/\drho_{b\Lambda},\brho_\Lambda]}
  \W_\Lambda[\brho_{b\Lambda}]=0.
\end{equation}
It should be noted that the $\brho_{b\Lambda}$-derivative hits all
$\brho_{b\Lambda}$ from the far left, that is, it acts on
$\brho_\Lambda$ in $\Delta S$ as well as on $\brho_{b\Lambda}$ in
$\W_\Lambda[\brho_{b\Lambda}]$.  A subtle point is that $\Delta S$
contains not only $\brho_{b\Lambda}$ but also $\brho_\Lambda$.  This
is because the origin of $\brho_\Lambda$ in $\Delta S$ is traced back
to the eikonal coupling $\sim \rho^a a^-_a$ for semi-fast gluons.  The
integration over $a^\mu_a$ leads to the renormalization
$\delta\brho_\Lambda$ for $\textit{soft}$ gluons, but not for
semi-fast gluons themselves.  It might seem then that taking the limit
of $b\to1$ in $\Delta S$ is a subtle business.  This does not matter
in fact, however, because there remains no contribution from the
$\brho_{b\Lambda}$-derivative hitting $\brho_{b\Lambda}$ inside
$\Delta S$ as we shall confirm explicitly below.  [We will see another
reasoning for that in App.~\ref{sec:conventional}.]

     The rest of this section would be devoted to illustrating the
evaluation of
$\Delta S[\A[\brho_{b\Lambda}]-i\delta/\drho_{b\Lambda},\brho_\Lambda]$.
At this point we realize that $\Delta S[\A[\brho_{b\Lambda}]-i\delta/
\drho_{b\Lambda},\brho_\Lambda]$ is the same quantity calculated in
Sec.~6 of \cite{Hatta:2005rn}, while (\ref{eq:delS}) was different.
We can borrow the results from \cite{Hatta:2005rn} but shall try to
make our calculations in a self-contained fashion.  We will work in
the light-cone gauge, which eventually leads to the same formula as
the calculation in the covariant gauge in \cite{Hatta:2005rn}, of
course.  By using $\partial^+\dA^-=0$ one can easily see
\begin{equation}
 \frac{\delta\Sym}{\dA^\mu_a}\biggr|_{\A[\brho_{b\Lambda}]
  -i\delta/\delta\brho_{b\Lambda}}\!\!-\delta^{\mu-}\brho_\Lambda
 =\delta^{\mu i}2ig\partial^+\Bigl\{ f^{abc}\frac{\delta}
  {\delta\brho_{b\Lambda}^b} \A^i_c[\brho_{b\Lambda}] \Bigr\}
  +\delta^{\mu-}\drho_\Lambda
\label{eq:dsda}
\end{equation}
in the light-cone gauge.  One can also express it in terms of the
quantities in the covariant gauge by means of
\begin{equation}
 f^{abc}\frac{\delta}{\delta\brho_{b\Lambda}^b}\A^i_c
  =\frac{2}{iN_c}\tr\Bigl[T^a\frac{\delta}{\delta\brho_{b\Lambda}}
  \A^i\Bigr]
  = \frac{1}{g}\,\Bigl(\partial^i\frac{\delta}{\delta\brho_{b\Lambda}^b}
  \Bigr)\,(V^\dagger_{x^-,-\infty})_{ba} .
\end{equation}
Here we substituted (\ref{eq:lc_sol}) and rotated the
$\brho_{b\Lambda}$-derivative by $V_{x^-,-\infty}$ to make it in the
covariant gauge and used a formula
$V_{ab}=(1/N_c)\tr[T^a V T^b V^\dagger]$.  Also we performed the
integration by parts to move $\partial^i$ onto
$\delta/\delta\brho_{b\Lambda}$.  Although we skip
the detailed explanations, we can use the \textit{free} propagator,
i.e.\ the propagator in the absence of the background field
$\A^i_a(\vec{x})$, if we employ the light-cone gauge.  This is,
roughly speaking, because in the light-cone gauge $\A^i_a(\vec{x})$
has no singularity but only a discontinuity at $x^-=0$ which does not
affect the propagator \cite{Ferreiro:2001qy}.  The free propagator can
be further approximated in the interested region as
\cite{Hatta:2005rn}
\begin{equation}
 G^{ij}_{0ab}(x,y) \simeq \delta^{ij}\delta^{ab}\frac{d\tau}{4\pi}\,
  \delta^{(2)}(\xt-\yt),
\label{eq:app_prop}
\end{equation}
where $d\tau=\ln[1/b]$.  The other components, $G^{i-}_{0ab}$,
$G^{-i}_{0ab}$, and $G^{--}_{0ab}$, are suppressed by $1/p^+$ as
compared with $G^{ij}_{0ab}$ and we drop them in the leading-log
approximation.  Therefore, under this approximation, we can neglect
$\delta^{\mu-}\drho_\Lambda$ in (\ref{eq:dsda}) because no logarithmic
enhancement results from the terms with $\delta^{\mu-}\drho_\Lambda$
in $\Delta S$.

     Because the first term in (\ref{eq:dsda}) is a total derivative
in $x^-$ and the approximate propagator (\ref{eq:app_prop}) does not
depend on $x^-$, the integrations over $x^-$ and $y^-$ are almost
trivial, so that only the surface terms at $x^-,y^-=\pm\infty$ remain.
For instance, the integration of the first term in (\ref{eq:dsda})
with respect to $x^-$ results in
\begin{equation}
 \delta^{\mu i}2i\biggl[\Bigl(\partial^i \frac{\delta}
  {\delta\brho_\tau^b(x^-_\tau,\xt)}\Bigr)V^\dagger_{ba}(\xt)
  -\Bigl(\partial^i\frac{\delta}{\delta\brho_\tau^a(0,\xt)}\Bigr)
  \biggr].
\end{equation}
In the above we have referred to the separation scale by using the
rapidity variable $\tau$ instead of $\Lambda$ or $b\Lambda$ and have
used the fact that $\brho_\tau(\vec{x})$ has a support
$0\lesssim x^-\lesssim x^-_\tau$.  The matrix $V$ is
$V_{x^-_\tau,-\infty}\simeq V_{+\infty,-\infty}$ with the definition
(\ref{eq:V}).  The integrand does not depend on time and thus the
$x^+$ and $y^+$ integrations are absorbed in the redefinition as
$\int\!dx^+\delta/\delta\brho_\tau(x)\to\delta/\delta\brho_\tau(\vec{x})$
(see (6.8) in \cite{Hatta:2005rn}).  After all $\Delta S$ takes a form
of
\begin{equation}
 \begin{split}
 \Delta S &=\frac{d\tau}{2\pi}\int\!d^2\xt \biggl[\Bigl(\partial^i
  \frac{\delta}{\delta\brho_\tau^a(x^-_\tau,\xt)}\Bigr)\Bigl(\partial^i
  \frac{\delta}{\delta\brho_\tau^a(x^-_\tau,\xt)}\Bigr) -\Bigl(\partial^i
  \frac{\delta}{\delta\brho_\tau^a(x^-_\tau,\xt)}\Bigr)\Bigl(\partial^i
  \frac{\delta}{\delta\brho_\tau^b(0,\xt)}\Bigr)V^\dagger_{ab}(\xt) \\
 &\qquad\qquad\qquad\qquad -\Bigl(\partial^i\frac{\delta}
  {\delta\brho^a(0,\xt)}\Bigr)\Bigl(\partial^i\frac{\delta}
  {\delta\brho^b(x^-_\tau,\xt)}\Bigr)V_{ab}(\xt)+\Bigl(\partial^i
  \frac{\delta}{\delta\brho^a(0,\xt)}\Bigr)\Bigl(\partial^i\frac{\delta}
  {\delta\brho^a(0,\xt)}\Bigr)\biggr],
 \end{split}
\label{eq:DS_result}
\end{equation}
which is nothing but (6.5) in \cite{Hatta:2005rn}.  It is apparent
from this expression that the $\brho_\tau$-derivative does not act on
$\brho_\tau$ in $\Delta S$ because
$\delta V^\dagger_{ab}/\delta\brho_\tau^a(x_\tau^-)\propto(T^a)_{ac}V^\dagger_{cb}=0$ and
$\delta V^\dagger_{ab}/\delta\brho_\tau^b(0)\propto V^\dagger_{ac}(T^b)_{cb}=0$,
etc.  Then in exactly the same way as in \cite{Hatta:2005rn}, we can
change the variable from $\brho_\tau^a(\vec{x})$ to
$\alpha_\tau^a(\vec{x})$ by (\ref{eq:poisson}) and with the help of
relations,
\begin{equation}
 \frac{\delta}{\delta\alpha_\tau^a(0,\xt)}
  =\frac{\delta}{\delta\alpha_\tau^b(x^-_\tau,\xt)}V^\dagger_{ba}(\xt)
  =V_{ab}(\xt)\frac{\delta}{\delta\alpha_\tau^b(x^-_\tau,\xt)},
\label{eq:relations}
\end{equation}
it is immediate to confirm that we finally reach the JIMWLK equation
(\ref{eq:JIMWLK}).  Let us focus on the second term in
(\ref{eq:DS_result}) to take an example.  Using (\ref{eq:poisson}),
(\ref{eq:eta}), and (\ref{eq:relations}) in order we have
\begin{align}
 &-\frac{d\tau}{2\pi}\int\!d^2\xt \Bigl(\partial^i\frac{\delta}
  {\drho_\tau^a(x_\tau^-,\xt)}\Bigr)\Bigr(\partial^i\frac{\delta}
  {\drho_\tau^b(0,\xt)}\Bigr) V^\dagger_{ab}(\xt) \notag\\
 = &-\frac{d\tau}{2\pi}\int\!d^2\xt d^2\yt d^2\zt \langle\zt|
  \frac{\partt}{\partt^2}|\xt\rangle \cdot \langle\zt|\frac{\partt}
  {\partt^2}|\yt \rangle \frac{\delta}{\delta
  \alpha_\tau^a(x_\tau^-,\xt)}\frac{\delta}{\delta\alpha_\tau^b(0,\yt)}
  V^\dagger_{ab}(\zt) \notag\\
 = &-\frac{d\tau}{(2\pi)^3}\int\!d^2\xt d^2\yt d^2\zt\,
  \K(\xt,\yt,\zt) \frac{\delta}{\delta\alpha_\tau^a(x_\tau^-,\xt)}
  V_{cb}(\yt)\frac{\delta}{\delta\alpha_\tau^b(x_\tau^-,\yt)}
  V^\dagger_{ac}(\zt),
\label{eq:example}
\end{align}
which is identical with the second term inside the square brackets in
the evolution kernel (\ref{eq:eta}).

     Before closing this section, we shall comment upon the meaning of
(\ref{eq:relations}).  We can easily validate these relations
supposing that it is only $V$ or $V^\dagger$ on which the
$\alpha_\tau$-derivatives act.  It does not mean, however, that
$\W_\tau[\brho_\tau]$ is a function of $V$ and $V^\dagger$.  In fact,
the derivatives can be moved from on $\W_\tau[\brho_\tau]$ to on
$\mathcal{O}[\brho_\tau]$ on averaging over $\alpha_\tau$ (or
$\brho_\tau$) in (\ref{eq:expectation}).  Thus, it is sufficient to
require $\mathcal{O}[\brho_\tau]$ written in terms of $V$ and
$V^\dagger$, which is actually the case if we consider the scattering
between the color glass condensate and the projectile consisting of
partons in the eikonal approximation.  From the practical point of
view we can say that the relations of (\ref{eq:relations}) would not
lose generality of the evolution equation.

     This treatment may well arise a question about the ordering
problem.  It is obvious from (\ref{eq:example}) that
$\delta V_{cb}(\yt)/\delta\alpha_\tau^a(\xt)$ is nonvanishing and so
$\delta/\delta\alpha_\tau^a(x_\tau^-)$ and
$\delta/\delta\alpha_\tau^b(0)$ do not commute after
(\ref{eq:relations}) is applied, while they should be commutative by
construction.  This is not a paradox, however.  For example, in
(\ref{eq:example}), there is certainly a non-zero contribution from
$\delta V_{cb}(\yt)/\delta\alpha_\tau^a(\xt)$, but such a term is
precisely compensated for by the term arising from non-commutativity
among $\alpha_\tau$-derivatives acting on $V$ or $V^\dagger$, that is,
\begin{equation}
 \biggl[ \frac{1}{i}\frac{\delta}{\delta\alpha_\tau^a(\xt)},\;
  \frac{1}{i}\frac{\delta}{\delta\alpha_\tau^b(\yt)}\biggr] = -igf^{abc}
  \frac{1}{i}\frac{\delta}{\delta\alpha_\tau^c(\xt)}\,\delta^{(2)}(\xt-\yt),
\label{eq:commut_deriv}
\end{equation}
if these derivatives are supposed to act on $V$ or $V^\dagger$.
Therefore, there is no ordering problem at all unless the quantities
in (\ref{eq:relations}) are split apart;  in (\ref{eq:relations}) an
ordering like
$(\delta/\delta\alpha_\tau^b(\yt))(\delta/\delta\alpha_\tau^a(\xt))V_{cb}(\yt)$
is not allowed, which generates extra terms and causes the ordering
ambiguity.  In other words, the ordering realizing in the JIMWLK
equation (\ref{eq:JIMWLK}) is determined uniquely in a way that
(\ref{eq:commut_deriv}) would bring about no ordering problem.

     In the present calculation we did not treat the virtual term
$\sigma^a$ (see (\ref{eq:virtual})) in an explicit manner.  The
virtual term is, in fact, included implicitly in the correct ordering
as pointed out in \cite{Hatta:2005rn} and it will be clearer in the
discussions in App.~\ref{sec:conventional}.


\section{conclusions}
\label{sec:conclusion}

     We have derived the JIMWLK equation in the functional integral
formalism.  Our starting point is different from the conventional
derivation;  we used a different representation of the QCD generating
functional consisting of the density of states and the simple eikonal
coupling.  In this way, we could clarify how the non-commutative
nature of quantum color charge density becomes irrelevant in the
approximation that the color charge density is large.  Actually the
expansion of the Wilson line in the non-local source term has turned
out to be responsible for loss of non-commutativity.  We further
discuss in App.~\ref{sec:conventional} some connections between the
approximations made in the new derivation and their counterparts in
the conventional derivation.  The comparison between two different
methods would give us a deeper understanding of those approximations.

     The next step would be how we can improve the approximation to
handle non-commutativity in the functional integral formalism.  The
commutation relation between the functional derivatives with respect
to the classical gauge field (see (\ref{eq:commut_deriv})) originates
from that the derivatives act on the gauge invariant Wilson lines, and
there is no difficulty in this case.  In contrast, in the case of the
commutation relation between the color charge density operators, the
functional integral description for that is non-trivial because the
color charge density in the functional integral is not an operator
like the derivative but a given classical field.  From our analysis we
would presume that the non-local source term written in terms of the
Wilson line takes care of this, that is, if one is capable to take
full account of the source term, one could reach a general evolution
equation in terms of classical quantities alone.  In our formalism
such a manipulation would be possible, at least in principle, by
diagonalizing the classical color source, and rotating it back in
color space.  We are making progress in this direction.  The point is,
once it will be accomplished, the quantization is automatic and there
is no need to impose the commutation relation between color charge
density like in \cite{Hatta:2005rn}.

     Although it is far from obvious whether the renormalization group
equation is closed in a color glass condensate formalism in an
intermediate region between the dense and dilute regimes, it would be
a challenging problem to attack it in the functional integral
formalism.  We believe that the present formulation would provide a
useful information for that purpose.

\acknowledgments

The author thanks L.~McLerran and Y.~Hatta for discussions.  He also
thanks E.~Iancu for a careful reading through the manuscript and
useful comments.  This work was supported by the RIKEN BNL Research
Center and the U.S.\ Department of Energy (D.O.E.) under cooperative
research agreement \#DE-AC02-98CH10886.

\appendix

\section{comparison to the conventional derivation}
\label{sec:conventional}

     In this paper we have elaborated the derivation of the JIMWLK
equation in the framework with the source terms consisting of the
density of states and the simple eikonal coupling.  Now, let us look
over the conventional derivation to grasp some connections between two
different strategies.

     The derivation we address here is not faithful to the original
one in literatures.  We shall rearrange the calculations and put our
emphasis on some aspects which are related to our derivation in the
text, though the main building blocks are not quite new.
The weight functional $\W_\Lambda[\brho_\Lambda]$ changes according to
$\Lambda$ in a way to keep the generating functional $Z$ intact, from
which the evolution equation follows.  That is, $Z$ can be written at
arbitrary separation scale $b\Lambda$ as in (\ref{eq:generating}).  We
shall take (Step-1), (Step-2), and (Step-3) respectively as in the
derivation we addressed previously. \vspace{5mm}

     (Step-1)\ \ The gauge fields are decomposed into the semi-fast
gluon $a^\mu_a$ with $b\Lambda<|p^+|<\Lambda$, the classical solution
$\A^i_a[\brho_\Lambda+\delta\rho]$, and the rest $\dA^\mu_a$ with
$|p^+|<b\Lambda$ , and then the action, $S=\Sym+S_W$, is expanded up
to the quadratic order in terms of $a^\mu_a$;
\begin{equation}
 \begin{split}
 \int^\Lambda\!\D A\,\delta[A^+]\,\exp\Bigl[iS[A,\brho_\Lambda]
  \Bigr] &\simeq \int^{b\Lambda}\!\!\D\dA\,\delta[\dA^+]
  \int_{b\Lambda}^\Lambda\!\!\D a\,\delta[a^+]\,\exp\Bigl[
  iS[\A+\dA,\brho_\Lambda]\Bigr] \\
 &\quad\times \exp\biggl[i\int\!d^4x\,\frac{\delta S}{\dA^\mu_a(x)}
  a^\mu_a(x)+\frac{i}{2}\int\!d^4x\,d^4y\,\frac{\delta^2 S}
  {\dA^\mu_a(x)\dA^\nu_b(y)} a^\mu_a(x) a^\nu_b(y) \biggr].
 \end{split}
\label{eq:act_exp}
\end{equation}
It should be mentioned that in this case the action contains
$S_W[A^-,\brho_\Lambda]$ given by (\ref{eq:source_log}) that generates
non-local terms in $x^+$.  This is an important difference from what
we have seen in our derivation in Sec.~\ref{sec:derivation}.

     In the dense regime where $\dA^\mu_a$ is small (\ref{eq:act_exp})
can be further expanded in terms of $\dA^\mu_a$ to read the quantum
corrections to the source as
\begin{equation}
 \int_{b\Lambda}^\Lambda\!\D a\,\delta[a^+]\,\exp\biggl[
  -\frac{1}{2}\int\!d^4x\,d^4y\,a^\mu_a(x)\,G^{-1ab}_{\mu\nu}(x,y)
  \,a^\nu_b(y) -i\int\!d^4x\,\delta\rho^a(x)\bigl(-a^-_a(x)+\dA^-_a(x)
  \bigr)\biggr],
\label{eq:exp_drho}
\end{equation}
where we have used the equations of motion including the source terms,
i.e.
\begin{equation}
 \frac{\delta S[A,\brho_\Lambda]}{\dA^\mu_a(x)}
  \biggr|_{\A[\brho_\Lambda+\delta\rho]} = \delta\rho^a(x)
\end{equation}
and we have defined the corrections to the source,
\begin{equation}
 \delta\rho^a(x)=\delta\rho^{(1)a}(x)+\delta\rho^{(2)a}(x) =
  -\int\!d^4y\,\frac{\delta^2 S}{\dA^-_a(x)\dA^\nu_b(y)}\biggr|_{\A}
  a^\nu_b(y) -\frac{1}{2}\int\!d^4y\,d^4z\,\frac{\delta^3 S}
  {\dA^-_a(x)\dA^\nu_b(y)\dA^\lambda_c(z)}\biggr|_{\A}a^\nu_b(y)
  a^\lambda_c(z).
\end{equation}
This approximation that one keeps only the linear term,
$-i\delta\rho^a(x)\dA^-_a(x)$, corresponds to the expansion of
(\ref{eq:exp_lnW}) and is responsible for loss of non-commutativity.
In writing (\ref{eq:exp_drho}) we have defined the propagator for
$a^\mu_a$ with $S_W$ included (that makes a difference from
$G^{-1ab}_{0\mu\nu}$ appearing in Sec.~\ref{sec:renormalization}) as
\begin{equation}
 iG^{-1ab}_{\mu\nu}[\brho_\Lambda](x,y) =\frac{\delta^2 S}
  {\dA^\mu_a(x)\dA^\nu_b(y)}\biggr|_{\A},
\label{eq:prop_withSw}
\end{equation}
where the argument $\brho_\Lambda$ of $G^{-1ab}_{\mu\nu}$ refers to
the source contained in $S_W$.
\vspace{3mm}

     (Step-2)\ \ The generating functional with the separation scale
$\Lambda$ is now rewritten in a form from which we can identify the
weight functional at scale $b\Lambda$, that is,
\begin{align}
 &\int\!\D\brho_\Lambda\,\W_\Lambda[\brho_\Lambda]\;
  \int^\Lambda\!\!\D\dA\,\delta[\dA^+]\,e^{iS[\A+\dA,\brho_\Lambda]}
  \notag\\
 =&\int\!\D\brho_\Lambda\,\W_\Lambda[\brho_\Lambda]\;
  \int_{b\Lambda}^\Lambda\!\!\D a\,\delta[a^+]\,e^{-\frac{1}{2}\int\!
  d^4x\,d^4y\,a^\mu_a(x)G^{-1ab}_{\mu\nu}(x,y)a^\nu_b(y) +i\int
  \!d^4x\,\delta\rho^a(x)a^-_a(x)} \int^{b\Lambda}\!\!\D\dA\,
  \delta[\dA^+]\,e^{iS[\A+\dA,\brho_\Lambda+\delta\rho]},
\end{align}
where the last term in (\ref{eq:exp_drho}) gives the renormalization
of the source $\brho_\Lambda\to\brho_\Lambda+\delta\rho$ in $S$ in the
above expression.  We shall introduce an auxiliary variable
$\brho_{b\Lambda}$ which is constrained as
$\brho_{b\Lambda}=\brho_\Lambda+\delta\rho$.  It should be worth
mentioning that $\delta\rho$ is a function of $a^\mu_a$ and should be
interpreted as the quantum color charge density in a sense in
discussions in Sec.~\ref{sec:quantum}.  In the dense approximation,
however, as we have explained before, $\dA^-$ is taken only up to the
first order and at that stage we have already lost the discrimination
between quantum and classical color charge density.  By introducing
$\brho_{b\Lambda}$ and integrating over $\brho_\Lambda$, we reach
\begin{align}
 &\int\!\D\brho_{b\Lambda} \int\!\D\brho_{\Lambda}\,\W_\Lambda
  [\brho_\Lambda]\; \int_{b\Lambda}^\Lambda\!\!\D a\,\delta[a^+]\,
  e^{-\frac{1}{2}\int\!d^4x\,d^4y\,a^\mu_a(x)G^{-1ab}_{\mu\nu}(x,y)
  a^\nu_b(y)+i\int\!d^4x\,\delta\rho^a(x)a^-_a(x)}\,
  \delta[\brho_{b\Lambda}-\brho_\Lambda-\delta\rho] \notag\\
 =&\int\!\D\brho_{b\Lambda}\int_{b\Lambda}^\Lambda\!\!\D a\,
  \delta[a^+]\,e^{-\frac{1}{2}\int\!d^4x\,d^4y\,a^\mu_a(x)
  G^{-1ab}_{\mu\nu}(x,y)a^\nu_b(y)+i\int\!d^4x\,\delta\rho^a(x)a^-_a(x)}\,
  \W_\Lambda[\brho_{b\Lambda}-\delta\rho],
\end{align}
where $\brho_\Lambda$ inside $G^{-1ab}_{\mu\nu}(x,y)$ must be altered
according to $\brho_\Lambda\to\brho_{b\Lambda}-\delta\rho$, though
this could be ignored in the Gaussian approximation with respect to
$a^\mu_a(x)$.  Otherwise, the already quadratic term
$-\frac{1}{2}a^\mu_a G^{-1ab}_{\mu\nu} a^\nu_b$ could produce higher
order terms.

     Here we shall remark that the role of the
$\delta\rho^a(x)a^-_a(x)$ term should deserve further investigations.
If we treat it in the Gaussian approximation, it cancels
$G^{-1ab}_{-\mu}$ and $G^{-1ab}_{\mu-}$, and is irrelevant to the
final results.  It has been anticipated in \cite{Hatta:2005wp},
however, that the complete integration over $a^\mu_a(x)$ including the
full Yang-Mills action would result in the Wess-Zumino term for
$\delta\rho^a(x)$.  In this paper, we shall just drop
$\delta\rho^a(x)a^-_a(x)$ hereafter.  This treatment is valid in
deriving the JIMWLK equation in the dense regime.

     Expanding the above expression in terms of $\delta\rho$ (with the
$\delta\rho^a(x)a^-_a(x)$ term dropped), we have
\begin{equation}
 \W_{b\Lambda}[\brho_{b\Lambda}]=\W_\Lambda[\brho_{b\Lambda}]
  -\int\!d^4x\,\frac{\delta}{\drho_{b\Lambda}^a(x)}\Bigl(\W_\Lambda
  [\brho_{b\Lambda}]\sigma^a(x)\Bigr) +\frac{1}{2}\int\!d^4x\,d^4y\,
  \frac{\delta^2}{\drho_{b\Lambda}^a(x)\drho_{b\Lambda}^b(y)}\Bigl(
  \W_\Lambda[\brho_{b\Lambda}]\chi^{ab}(x,y)\Bigr) ,
\label{eq:evolv_w}
\end{equation}
where the correlation functions are defined as
\begin{align}
 \sigma^a(x) &=\langle\delta\rho^{(2)a}(x)\rangle
  =-\frac{1}{2}\int\!d^4y\,d^4z\,\frac{\delta^3 S}{\dA^-_a(x)
  \dA^\nu_b(y)\dA^\lambda_c(z)}\cdot G^{\nu\lambda}_{bc}(y,z)
\label{eq:virtual} \\
 \chi^{ab}(x,y) &=\langle\delta\rho^{(1)a}(x)\delta\rho^{(1)b}(y)\rangle
  =\int\!d^4y\,d^4z\,d^4w\,\frac{\delta^2 S}{\dA^-_a(x)\dA^\lambda_c(z)}
  \cdot\frac{\delta^2 S}{\dA^-_b(y)\dA^\sigma_d(w)}\cdot
  G^{\lambda\sigma}_{cd}(z,w)
\end{align}
within the Gaussian approximation.  This is exactly the same
expression as (3.53) in \cite{Iancu:2000hn} and the evaluation of
these correlation functions goes just in the same manner.  We shall
comment that in this case the determinant as a result of the Gaussian
integration would not lead to the determinant (\ref{eq:renorm})
because the propagator defined in this section does not contain
$\dA^-$ in it.

     As we have already mentioned, we could have ignored the
difference by $\delta\rho$ inside the propagator within the Gaussian
approximation.  Then the $\brho_{b\Lambda}$-derivatives in
(\ref{eq:evolv_w}) would not hit the $\brho_\Lambda$ dependence in the
propagator.  This fact seems to correspond to the arguments in
Sec.~\ref{sec:renormalization} that the $\brho_{b\Lambda}$-derivatives
vanish acting on $\brho_{b\Lambda}$ in $\Delta S$.  We cannot see the
mathematical correspondence transparently, however, for the propagator
(\ref{eq:prop_withSw}) involves non-local contributions from
$S_W[A^-,\brho]$.

     It is known that the explicit evaluation reveals an interesting
relation,
\begin{equation}
 \frac{1}{2}\int\!d^4y\,\frac{\delta\chi^{ab}(x,y)}
  {\delta\brho_{b\Lambda}^b(y)} = \sigma^a(x),
\label{eq:sigma_chi}
\end{equation}
from which (\ref{eq:evolv_w}) simplifies as
\begin{equation}
 \frac{\partial}{\partial\tau}\W_\tau[\brho_\tau]
  =\frac{1}{2}\int\!d^4x\,d^4y\,\frac{\delta}{\drho_\tau^a(x)}
  \chi^{ab}(x,y)\frac{\delta}{\drho_\tau^b(y)}\W_\tau[\brho_\tau].
\label{eq:evolv_w_simple}
\end{equation}
This corresponds to the fact, as we mentioned before in
Sec.~\ref{sec:renormalization}, the correct ordering implies the
virtual term $\sigma^a(x)$.  We obtained in
Sec.~\ref{sec:renormalization} the same ordering as in
(\ref{eq:evolv_w_simple}) which is determined by the condition that
the commutation relation (\ref{eq:commut_deriv}) becomes free from the
ordering problem.   The ordering as shown in (\ref{eq:evolv_w_simple})
implicitly signifies the presence of the virtual contribution through
(\ref{eq:sigma_chi}).  In this way, we should consider that our
derivation contains the virtual term as well in an implicit way, which
has also been mentioned in part in \cite{Hatta:2005rn}.



\begin{thebibliography}{99}

\bibitem{Lipatov:1976zz}
  L.~N.~Lipatov,
  Sov.\ J.\ Nucl.\ Phys.\  {\bf 23}, 338 (1976)
  [Yad.\ Fiz.\  {\bf 23}, 642 (1976)].

\bibitem{Kuraev:1976ge}
  E.~A.~Kuraev, L.~N.~Lipatov and V.~S.~Fadin,
  Sov.\ Phys.\ JETP {\bf 44}, 443 (1976)
  [Zh.\ Eksp.\ Teor.\ Fiz.\  {\bf 71}, 840 (1976)].

\bibitem{Kuraev:1977fs}
  E.~A.~Kuraev, L.~N.~Lipatov and V.~S.~Fadin,
  Sov.\ Phys.\ JETP {\bf 45}, 199 (1977)
  [Zh.\ Eksp.\ Teor.\ Fiz.\  {\bf 72}, 377 (1977)].

\bibitem{Balitsky:1978ic}
  I.~I.~Balitsky and L.~N.~Lipatov,
  Sov.\ J.\ Nucl.\ Phys.\  {\bf 28}, 822 (1978)
  [Yad.\ Fiz.\  {\bf 28}, 1597 (1978)].

\bibitem{Gribov:1984tu}
  L.~V.~Gribov, E.~M.~Levin and M.~G.~Ryskin,
  Phys.\ Rept.\  {\bf 100}, 1 (1983).

\bibitem{Mueller:1988xy}
  A.~H.~Mueller,
  Nucl.\ Phys.\ B {\bf 307}, 34 (1988).

\bibitem{McLerran:1993ni}
  L.~D.~McLerran and R.~Venugopalan,
  Phys.\ Rev.\ D {\bf 49}, 2233 (1994)
  [arXiv:hep-ph/9309289].

\bibitem{McLerran:1993ka}
  L.~D.~McLerran and R.~Venugopalan,
  Phys.\ Rev.\ D {\bf 49}, 3352 (1994)
  [arXiv:hep-ph/9311205].

\bibitem{McLerran:1994vd}
  L.~D.~McLerran and R.~Venugopalan,
  Phys.\ Rev.\ D {\bf 50}, 2225 (1994)
  [arXiv:hep-ph/9402335].

\bibitem{review}
  For reviews on the color glass condensate, see;
  E.~Iancu, A.~Leonidov and L.~McLerran,
  arXiv:hep-ph/0202270;
  E.~Iancu and R.~Venugopalan,
  arXiv:hep-ph/0303204.

\bibitem{Golec-Biernat:2001if}
  K.~Golec-Biernat, L.~Motyka and A.~M.~Stasto,
  Phys.\ Rev.\ D {\bf 65}, 074037 (2002)
  [arXiv:hep-ph/0110325].

\bibitem{Jalilian-Marian:1997jx}
  J.~Jalilian-Marian, A.~Kovner, A.~Leonidov and H.~Weigert,
  Nucl.\ Phys.\ B {\bf 504}, 415 (1997)
  [arXiv:hep-ph/9701284].

\bibitem{Jalilian-Marian:1997gr}
  J.~Jalilian-Marian, A.~Kovner, A.~Leonidov and H.~Weigert,
  Phys.\ Rev.\ D {\bf 59}, 014014 (1999)
  [arXiv:hep-ph/9706377].

\bibitem{Jalilian-Marian:1997dw}
  J.~Jalilian-Marian, A.~Kovner and H.~Weigert,
  Phys.\ Rev.\ D {\bf 59}, 014015 (1999)
  [arXiv:hep-ph/9709432].

\bibitem{Kovner:2000pt}
  A.~Kovner, J.~G.~Milhano and H.~Weigert,
  Phys.\ Rev.\ D {\bf 62}, 114005 (2000)
  [arXiv:hep-ph/0004014].

\bibitem{Iancu:2000hn}
  E.~Iancu, A.~Leonidov and L.~D.~McLerran,
  Nucl.\ Phys.\ A {\bf 692}, 583 (2001)
  [arXiv:hep-ph/0011241].

\bibitem{Iancu:2001ad}
  E.~Iancu, A.~Leonidov and L.~D.~McLerran,
  Phys.\ Lett.\ B {\bf 510}, 133 (2001)
  [arXiv:hep-ph/0102009].

\bibitem{Ferreiro:2001qy}
  E.~Ferreiro, E.~Iancu, A.~Leonidov and L.~McLerran,
  Nucl.\ Phys.\ A {\bf 703}, 489 (2002)
  [arXiv:hep-ph/0109115].

\bibitem{Fukushima:2005kk}
  K.~Fukushima,
  arXiv:hep-ph/0512138 to appear in Nucl.\ Phys.\ A.

\bibitem{Blaizot:2002xy}
  J.~P.~Blaizot, E.~Iancu and H.~Weigert,
  Nucl.\ Phys.\ A {\bf 713}, 441 (2003)
  [arXiv:hep-ph/0206279].

\bibitem{Jalilian-Marian:2000ad}
  J.~Jalilian-Marian, S.~Jeon and R.~Venugopalan,
  Phys.\ Rev.\ D {\bf 63}, 036004 (2001)
  [arXiv:hep-ph/0003070].

\bibitem{Dirac}
  P.~A.~M.~Dirac, \textit{Lectures in Quantum Mechanics},
   Belfer Graduate School of Science, Yeshiva University Press,
   New York, 1964.

\bibitem{Hatta:2005wp}
  Y.~Hatta,
  arXiv:hep-ph/0511287.

\bibitem{Mueller:2005ut}
  A.~H.~Mueller, A.~I.~Shoshi and S.~M.~H.~Wong,
  Nucl.\ Phys.\ B {\bf 715}, 440 (2005)
  [arXiv:hep-ph/0501088].

\bibitem{Kovner:2005nq}
  A.~Kovner and M.~Lublinsky,
  Phys.\ Rev.\ D {\bf 71}, 085004 (2005)
  [arXiv:hep-ph/0501198].

\bibitem{Hatta:2005rn}
  Y.~Hatta, E.~Iancu, L.~McLerran, A.~Stasto and D.~N.~Triantafyllopoulos,
  Nucl.\ Phys.\ A {\bf 764}, 423 (2006)
  [arXiv:hep-ph/0504182].

\bibitem{Mueller:1993rr}
  A.~H.~Mueller,
  Nucl.\ Phys.\ B {\bf 415}, 373 (1994).

\bibitem{Iancu:2003uh}
  E.~Iancu and A.~H.~Mueller,
  Nucl.\ Phys.\ A {\bf 730}, 460 (2004)
  [arXiv:hep-ph/0308315].




\end{thebibliography}
\end{document}